\documentclass[12pt]{iopart}
\usepackage{graphicx}
\begin{document}

\title[\textit{Ab initio} investigation of impurity ferromagnetism in the Pd$_{1-x}$Fe$_{x}$ alloys]{\textit{Ab initio} investigation of impurity ferromagnetism in the Pd$_{1-x}$Fe$_{x}$ alloys:  concentration and position dependence}

\author{I.~I.~Piyanzina$^{1,2}$, A.~I.~Gumarov$^{1,2}$,  R.V. Yusupov$^{2}$,
 R.~I.~Khaibullin$^{1}$ and L.~R.~Tagirov$^{1,2}$}

\address{$^{1}$Zavoisky Physical-Technical Institute, FRC Kazan Scientific Center of RAS, 420029 Kazan, Russia}
	\address{$^{2}$Institute of Physics, Kazan Federal University,
             420008 Kazan, Russia}
     %   \author{A.~I.~Gumarov}

%\address{Zavoisky Physical-Technical Institute, FIC KazanSC of	RAS, 420029 Kazan, Russia}
%	\address{Institute of Physics, Kazan Federal University,
%             420008 Kazan, Russia}     
\ead{i.piyanzina@gmail.com}
\vspace{10pt}
\begin{indented}
\item[]May 2021
\end{indented}

\begin{abstract}
We present the \textit{ab initio} results of the structural and magnetic properties of the Pd host matrix implanted with Fe solute atoms %metal ions of Mn$^{+}$, Fe$^{+}$, Co$^{+}$, Ni$^{+}$, as well as  Gd$^{+}$, Er$^{+}$, Yb$^{+}$ 
at various concentrations. By means of density functional theory we confirm that iron impurities are able to initialize significant magnetization of the Pd atoms, when the impurity consentation exceeds 3\,at.$\%$. Besides, we demonstrate that the imposed magnetization depends on impurity positions in the host matrix, in particular, there is a maximum of magnetization for a uniform distribution of the iron impurity. %Finally, we compare the results for Fe$^{+}$ ions with other magnetic impurities in terms of imposed magnetization of the Pd host matrix.

\end{abstract}

\noindent{\it Keywords}: Pd-Fe alloy, impurity ferromagnetism,  DFT
%\maketitle

\section{Introduction}
The Pd$_{1-x}$Fe$_{x}$ alloys have been known and studied for a long time ago, pioneering works date back to 1938 by Fallot~\cite{fallot}, and later, in 1960, bulk Pd$_{1-x}$Fe$_{x}$ alloys were deeply investigated by Crangle in Ref.~\cite{crangle}. Experimental findings and their understanding at the time of the 70s were summarized in a review by Nieuwenhuys~\cite{nieu}. 
The current revival of interest has been fueled by a potential use as a material of a choice for a weak and soft ferromagnet in superconducting Josephson magnetic random-access memory (MRAM) based on Josephson junctions ~\cite{ryazanov,arham,bolginov,larkin,ryazanov2,vernik,niedziel,gingrich,glick,soloviev,usp}. The Pd-rich ferromagnetic  Pd$_{1-x}$Fe$_{x}$ ($0.01 < x < 0.1$) alloy is used there in a form of a thin film fabricated utilizing magnetron sputtering ~\cite{arham,bolginov,larkin,ryazanov2,vernik,niedziel,gingrich,glick,usp,uspenskaya1,bolginov2,uspenskaya2}, molecular-beam epitaxy~\cite{bolginov2,uspenskaya2,garifullin,ewerlin,esmaeili,esmaeili2,petrov,esmaeili3,mohammed,ya}, and ion-beam implantation~\cite{gumarov}. 

There is a scatter of opinions on compositional and magnetic homogeneity of Pd$_{1-x}$Fe$_{x}$ films obtained by different deposition techniques: some papers report on an indication of clustering ~\cite{niedziel,gingrich,glick,usp,uspenskaya1} or formation of the nanograins of Pd$_{3}$Fe phase~\cite{bolginov2,uspenskaya2} in samples deposited with the magnetron sputtering technique; the others on MBE growth of thin Pd$_{1-x}$Fe$_{x}$ films~\cite{esmaeili,esmaeili2,petrov,esmaeili3,mohammed,ya} report on their high magnetic homogeneity; finally, the ion-implanted palladium films~\cite{gumarov} show a significant influence of the Fe implant distribution on saturation magnetization ($M_{s}$) and Curie temperature ($T_{C}$) of the samples. It is not clear why and how the redistribution of iron solute in the Pd matrix influences $M_{s}$ and $T_{C}$. 
From very few existing studies of the electronic structure of Pd$_{1-x}$Fe$_{x}$ solid solutions, the Korrringa-Kohn-Rostoker Green’s function method with the local-density functional approximation was used to caltulate the magnetic moment at the palladium site in the dilute substitutions ($\textit{x}$ close to 1.0) of Pd in bcc Fe~\cite{opahle}, while the electronic structures and magnetic properties of Pd$_{1-x}$Fe$_{x}$ alloys with $0.5\leq x\leq 0.85$ were investigated in the framework of density functional theory (DFT) using the full potential approximation~\cite{drittler}. The only in a Pd-rich side is Ref.~\cite{burzo}, where the magnetic properties of palladium-iron alloys and compounds were calculated by means of spin-polarized and scalar relativistic tight-binding linear muffin tin orbital method (TB-LMTO) within atomic sphere approximation (ASA), together with the coherent potential approximation (CPA) to describe random Pd-Fe solutions. However, magnetic moments and Curie temperatures at $x < 0.1$ were not presented, therefore, the calculations of Ref.~\cite{burzo} are not applicable to Pd-rich alloys.

In this work, we perform $\textit{ab initio}$ calculations for the Pd$_{1-x}$Fe$_{x}$ alloy with the iron dopant uniformly distributed over  the bulk Pd host lattice at the different iron content $x < 0.1$ and calculate the mean magnetic moment per Fe solute atom and the maximal magnetic moment located on a Pd atom as a function of $x$. Then, we study the influence of inhomogeneity in the iron dopant distribution on the magnetic moment per Fe atom and the size and shape of the magnetized “bubble” of the host Pd atoms around the Fe aggregates.
% Finally, we  present results for other impurities (Co$^{+}$, Mn$^{+}$, Ni$^{+}$, as well as Yb$^{+}$, Gd$^{+}$, Er$^{+}$) for the comparison

%In order to answer to the question, 
%we will perform  $\textit{ab initio}$ calculations for Fe$^{+}$ ions located in the bulk Pd host matrix in different concentrations in order to correlate old experimental results~\cite{fallot,crangle} and our resent experimental investigations~\cite{esmaeili3,gumarov}. In order to answer to the above-mentioned question, we will also investigate magnetization versus impurity positions in the host matrix. Finally, we will present results for other impurities (Co$^{+}$, Mn$^{+}$, Ni$^{+}$, as well as Yb$^{+}$, Gd$^{+}$, Er$^{+}$) for the comparison.

\section{Computational details}
\begin{figure}
\center
\includegraphics[angle=0,width=6cm]{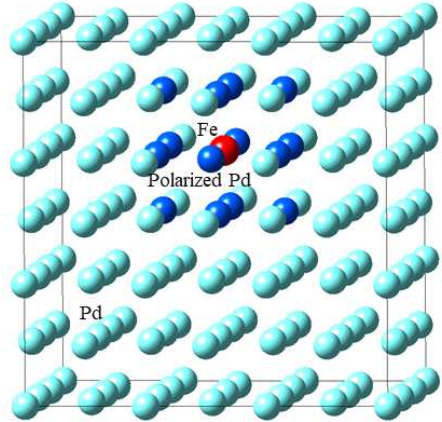}
	\caption{Supercell of Pd$_{0.98}$Fe$_{0.02}$ matrix used in our calculations  with denoted polarized Pd atoms around the Fe solute ion after the optimization procedure. Red sphere of atom corresponds to Fe, light blue -- to Pd, blue -- to predominantly polarized Pd atoms.}
	\label{cell}
\end{figure}

Our $\textit{ab initio}$ investigations were based on the DFT~\cite{hohenberg1964,kohn1965} approach within the VASP code~\cite{kresse1996a,kresse1996b,kresse1999} as a part of the MedeA\textsuperscript{\textregistered} software of Materials Design~\cite{medea}. Exchange and correlation effects were accounted for by the generalized gradient approximation (GGA)  as parameterized by Perdew, Burke, and Ernzerhof (PBE)~\cite{perdew1996}. The Kohn-Sham equations were solved using the plane-wave basis set (PAW)~\cite{bloechl1994paw}. The cut-off energy was chosen equal to 400\,eV. The force tolerance was 0.5\,eV/nm and the energy tolerance for the self-consistency loop was 10$^{-5}$\,eV. The Brillouin zones were sampled sampled using Monkhorst-Pack grids~\cite{monkhorst1976} including 3$\times$3$\times$3\,$\textbf{k}$-points. We performed spin-polarized calculations in all cases initializing Fe atoms to have 3.63%, 2.4, 2.7, 1.26, 9.5, 4.6 and 8
\,$\mu_{B}$,  and Pd atoms to be in the paramagnetic state (0\,$\mu_{B}$). %Finally, the electronic densities of states were calculated using the linear tetrahedron method~\cite{bloechl1994ltm} on $ 4 \times 4 \times 1 $ $ {\bf k} $-point grids.
The structures are described as consisting of a filled FCC host matrix formed by Pd atoms, with Fe ions substituting octahedrally coordinated sites only (see Fig.~\ref{cell}). The 3$\times$3$\times$3  unit cell parameter $a=1.18146$\,nm.
%Fig.1

\section{Results}
\subsection{Magnetic properties of Pd$_{1-x}$Fe$_{x}$ with varying concentration and uniform impurity distribution}

\begin{figure}
	\center
	\includegraphics[angle=0,width=8 cm]{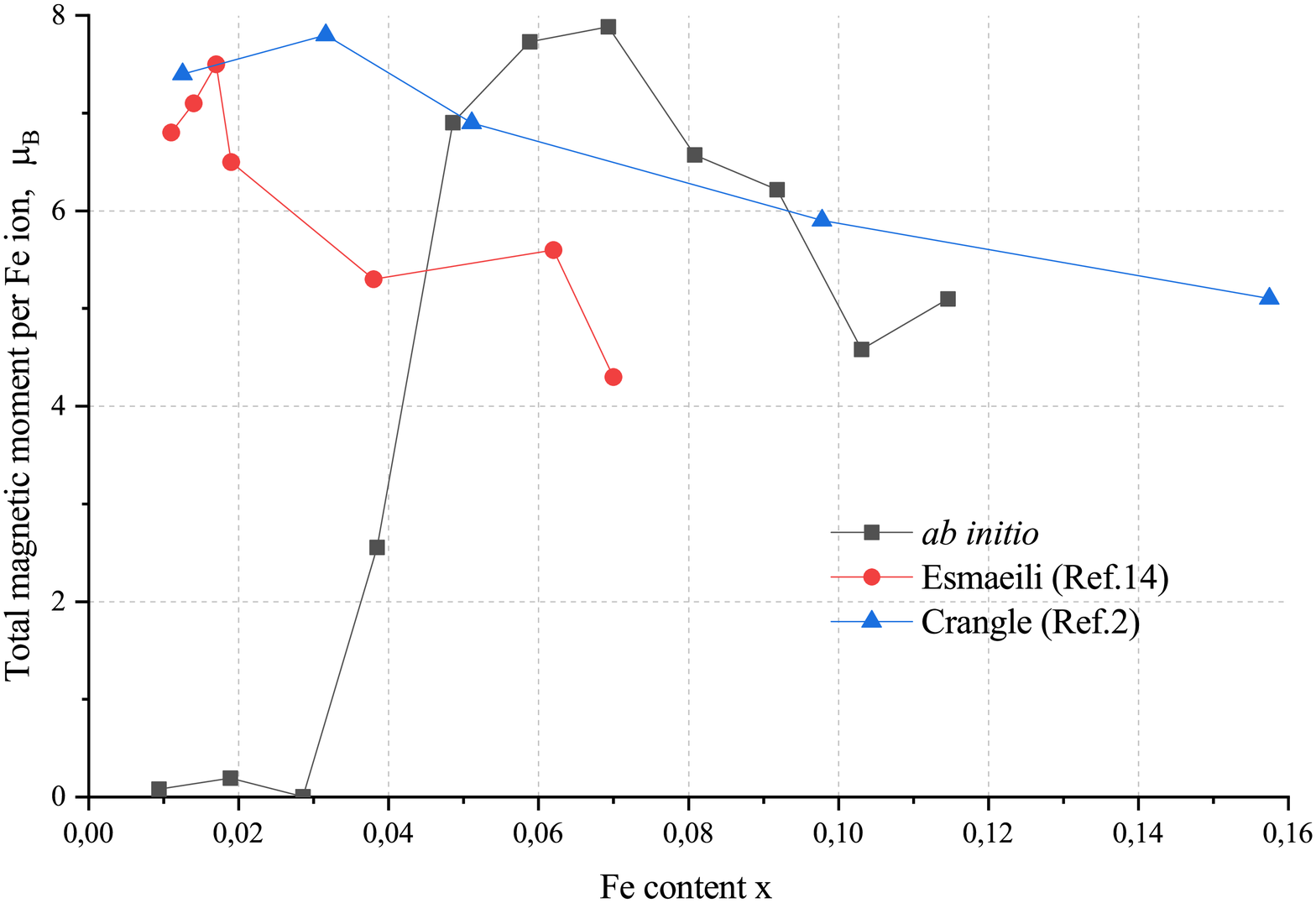} a)
	\includegraphics[angle=0,width=6.5 cm]{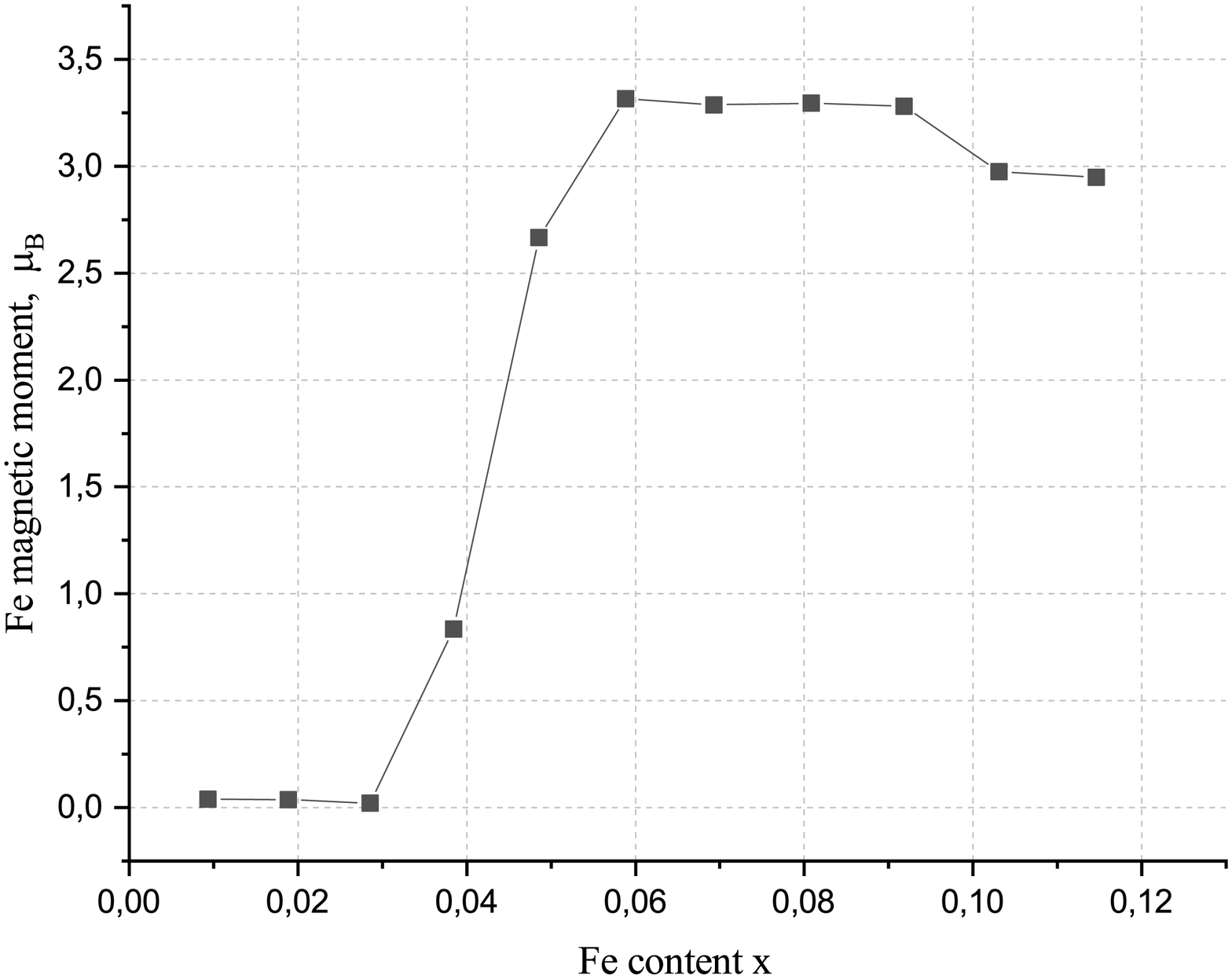} b)
	\caption{(a) Total calculated magnetic moment calculated per Fe solute atom \textit{versus} Fe content in the 3$\times$3$\times$3 Pd-supercell (black line with squares), along with the experimental data plotted for  MBE film (red line with circles) and for the bulk  (blue line with triangles)   taken from references \cite{esmaeili3} and \cite{crangle}, respectively.
		(b) The calculated average magnetic moment of Fe atom. 
		\newline
		The uniform impurity distribution is considered.}
	\label{Fe_1}
\end{figure}

\begin{figure}
	\center
	\includegraphics[angle=0,width=8 cm]{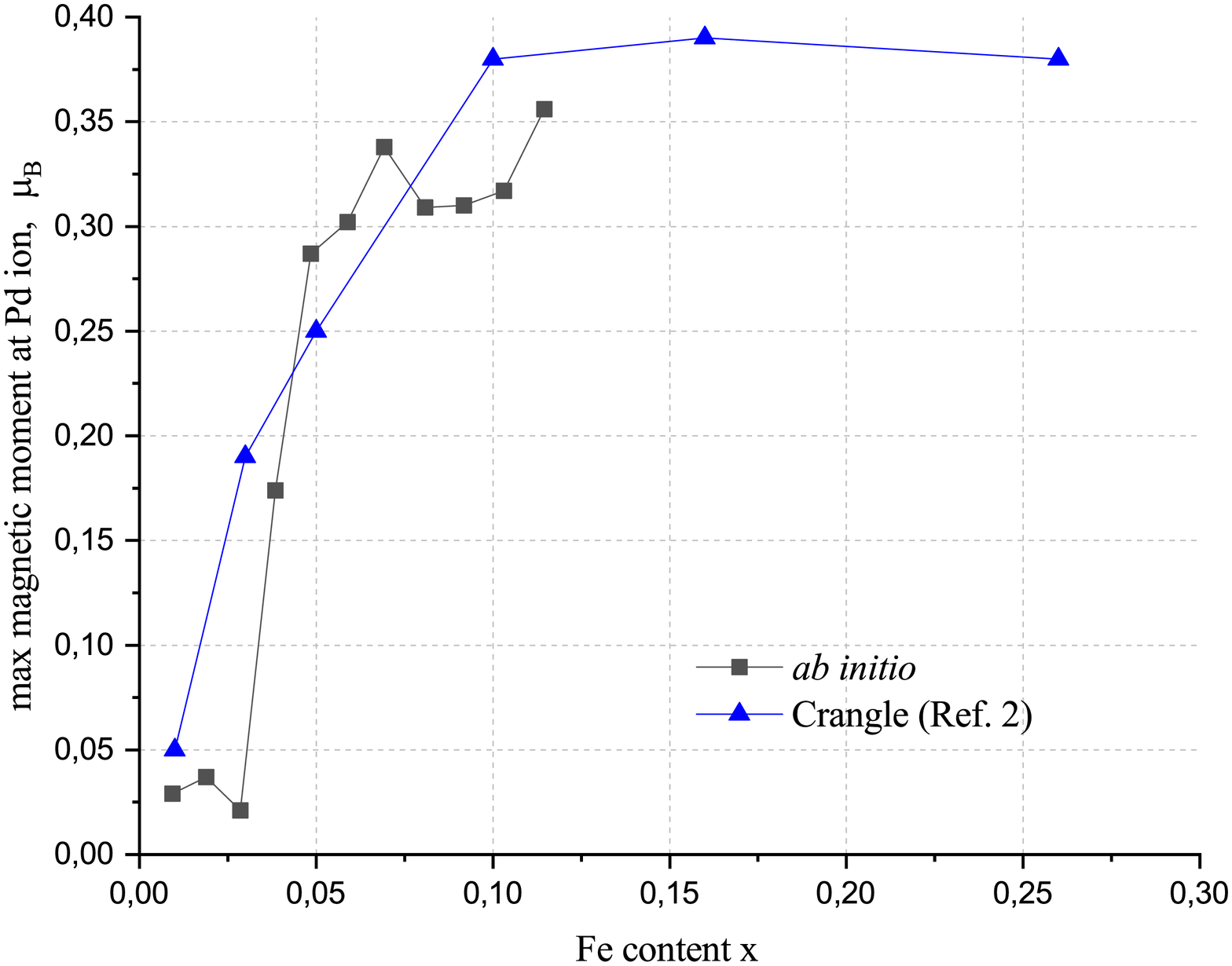} a)
	\includegraphics[angle=0,width=6.5 cm]{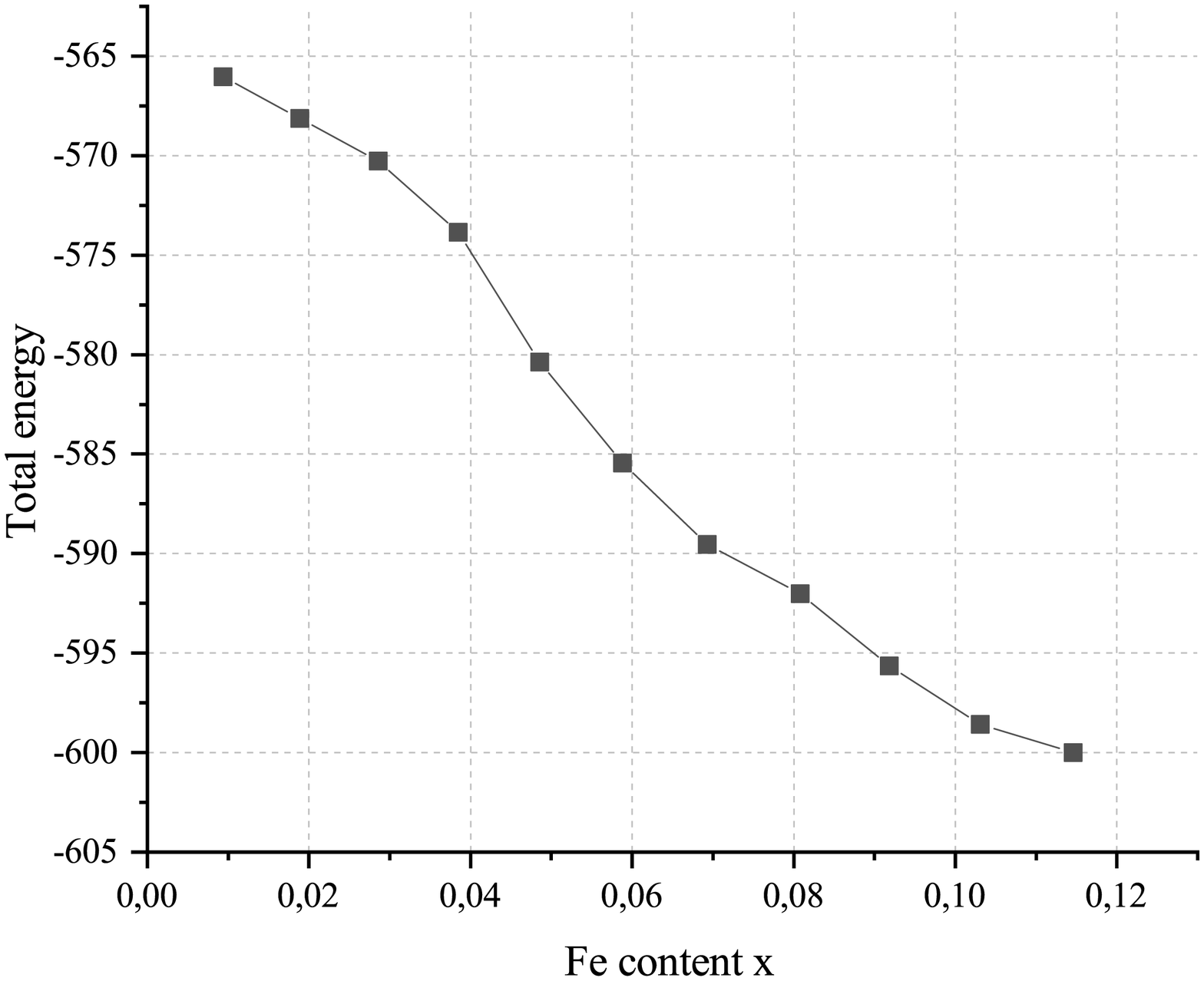} b)
	\caption{(a) Maximal magnetic moment calculated for a Pd atom \textit{versus} the Fe$^{+}$ atoms content in a 3$\times$3$\times$3 Pd-supercell with uniform impurity distribution (black line with squares) along with the experimental data for the bulk alloy~\cite{crangle} (blue line with triangles). 
		\newline
		(b) Energy decrease with Fe atoms content \textit{x} increasement. Here, total energy is calculated per 3$\times$3$\times$3 Pd-supercell with uniform distribution of Fe solute atoms.}
	\label{Fe_2}
\end{figure}

At the first stage, various concentrations of Fe solute in a 3$\times$3$\times$3 Pd-supercell with uniform distribution were considered. We increased the number of Fe atoms in the unit supercell from one to eleven by substituting Pd ions.
At low Fe$^{+}$ contents ($x=0.01$ to 0.03) we obtained a negligible cell magnetization (Fig.~\ref{Fe_1}\,a). At such concentrations, Fe solute atoms are non-magnetic (Fig.~\ref{Fe_1}\,b) and do not magnetize the surrounding Pd atoms. 
As the impurity concentration increases, iron becomes magnetic (Fig.~\ref{Fe_1}\,b) and magnetizes the surrounding Pd atoms (Fig.~\ref{Fe_2}\,a). In particular, about fourteen Pd atoms received significant magnetic moments, as shown in Fig.~\ref{cell}\,b. 
We found that the total magnetization of a cell with seven Fe solute atoms gives rise to the highest total magnetic moment calculated per Fe atom which equals $\approx$8\,$\mu_{B}$ (including Fe magnetic moment). This value is in agreement with the experimental results shown also in Fig.~\ref{Fe_1}.
After the peak point, the curve slowly decreases in agreement with results of Esmaeili \textit{et al.}~\cite{esmaeili3} obtained for MBE films, and of Crangle~\cite{crangle} found for the bulk. Our curve is somewhere in the middle of two mentioned results.
However, there is a contradiction with experiment at low impurity concentrations. That might be due to the fact that in the calculations were performed assuming 0\,K and we got Fe solute atoms, as a result of optimization, to be non-magnetic at low impurity concentrations, whereas experimental measurements were carried out at final temperatures. Besides, the theory of impurity ferromagnetism suggests that at very low impurity concentration, when the distance between impurity ions is large enough, the oscillatory potential prevails over the ferromagnetic one, so that the impurity ferromagnetism does not occur and the spin glass is formed in the alloy~\cite{korenblit}. In comparison with assumed theoretical value of $x=10^{-4}$~\cite{korenblit} for the critical Fe content, \textit{ab initio} gives $x=0.03$.
%Fig.2

At the same time, the calculated magnetic moments for Pd atoms are in a very good agreement with the experimental data obtained by Crangle~\cite{crangle} for bulk alloys (Fig.~\ref{Fe_2}\,a). One can notice that the magnetic moment of Pd sharply increases from $\approx 0.05$ to 0.35\,$\mu_{B}$  when the iron concentration exceeds three atoms per 3$\times$3$\times$3 cell. 
After that, the magnetic moment reaches a plateau with the highest value of the moment per Pd atom  equal to $\approx 0.35\,\mu_{B}$, which is consistent  with the experimental data  shown also in Fig.~\ref{Fe_2}\a. The magnetic moment of iron also reaches a plateau with a constant value of $\approx$3.25\,$\mu_{B}$ (Fig.~\ref{Fe_1}\,b), which is lower than the theoretical maximum, but it is higher than the value of 2.8\,$\mu_{B}$ obtained in Ref.~\cite{crangle} and close to the value of 3.5\,$\mu_{B}$ obtained by Neutron diffraction experiments~\cite{low,aldred}.
We also checked the concentration dependence of the total energy, which shows a descending character and might mean that the system tends to have Fe ions in the matrix (Fig.~\ref{Fe_2}\,b).
%Fig.3

The discussed findings confirm the reproducibility of experimental results~\cite{crangle,esmaeili3}  as well as theoretical predictions~\cite{korenblit} for Pd$_{1-x}$Fe$_{x}$ alloys with the $\textit{ab initio}$ instrument, which will be used in further discussions regarding the positions of Fe impurities in the host matrix.

\subsection{Pd$_{0.98}$Fe$_{0.02}$: impurity position impact}

\begin{figure}
\center
\includegraphics[angle=0,width=4.5cm]{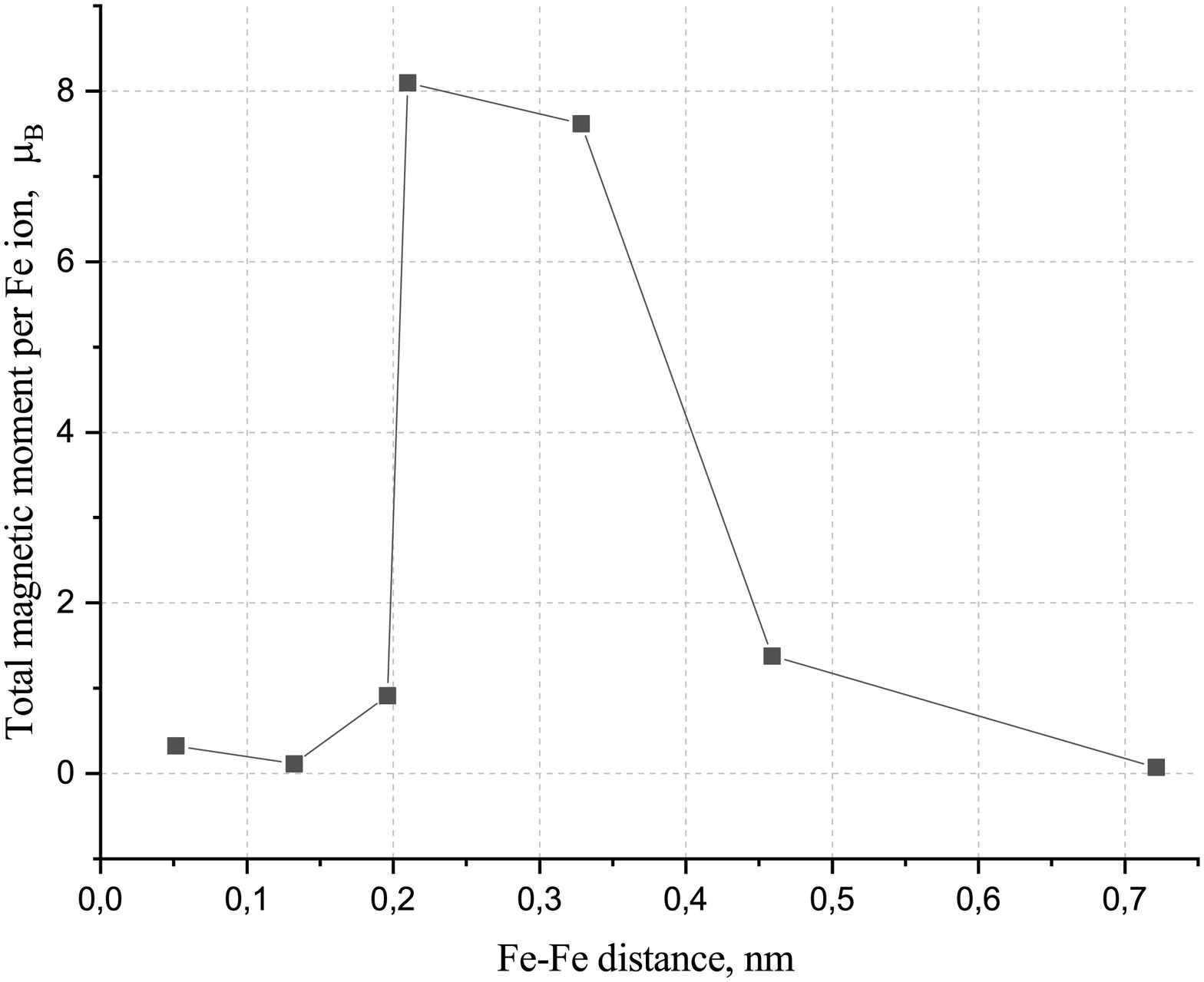} a)
\includegraphics[angle=0,width=4.5cm]{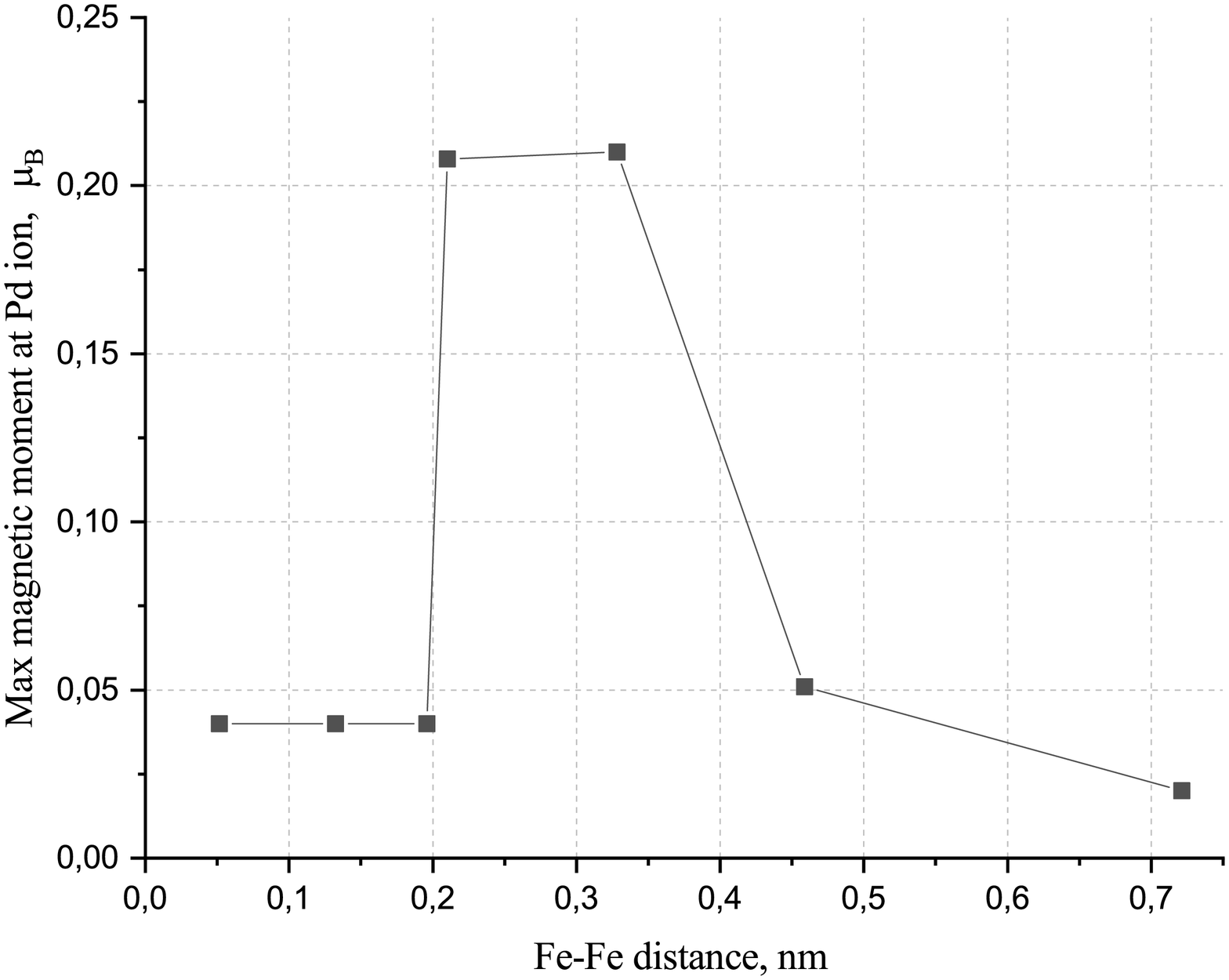} b)
\includegraphics[angle=0,width=4.5cm]{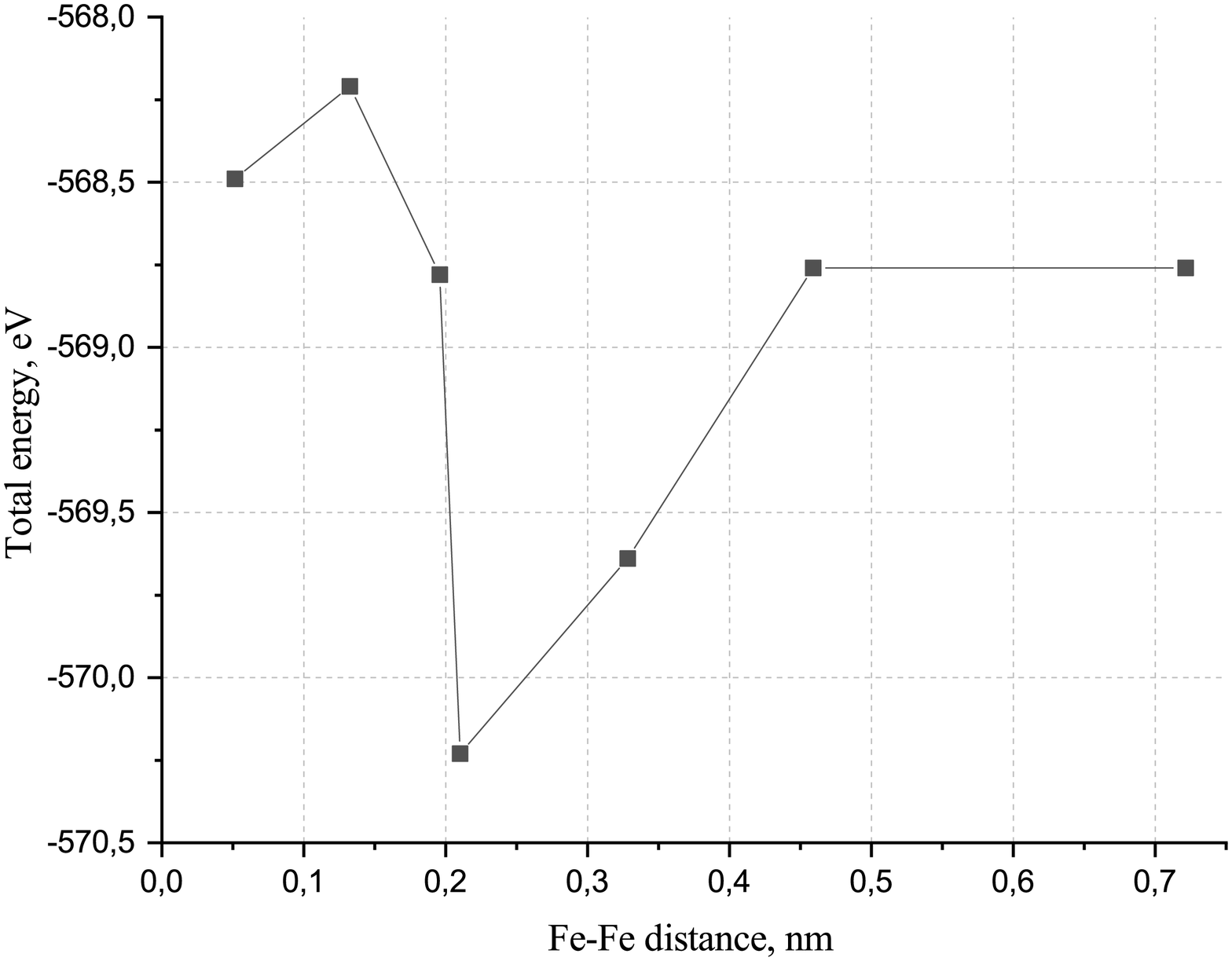} c)
	\caption{(a) Total magnetic moment calculated per Fe$^{+}$ ion plotted against Fe to Fe distance, (b) maximal magnetic moment at Pd atom,  and (c) total energy of the  3$\times$3$\times$3 Pd cell. Two substitutional Fe atoms in the host matrix cell were considered.}
	\label{vs_position}
\end{figure}

At the second stage, we consider the influence of Fe-Fe distance in the alloy with Pd$_{0.98}$Fe$_{0.02}$ composition if two iron solute atoms reside in the  3$\times$3$\times$3 Pd-supercell. 
We found that at particular distances ($\approx$\,0.2-0.35\,nm) the total magnetization as well as the magnetic moment per Pd atom reach a wide maximum (Fig.~\ref{vs_position}\,a). The highest total magnetization per Fe atom equals $\approx$\,8\,$\mu_{B}$, while the highest magnetic moment of the Pd atom is $\approx$0.2\,$\mu_{B}$ (Fig.~\ref{vs_position}\,b). This maximum of magnetization corresponds to the minimum of the total energy (Fig.~\ref{vs_position}\,c). We also tested a different (another concentration) configuration of three Fe solute atoms in the host Pd supercell and obtained similar dependence with a maximum of magnetization at analogous Fe-Fe distances.
 %Fig 4

\begin{figure}
\center
\includegraphics[angle=0,width=15cm]{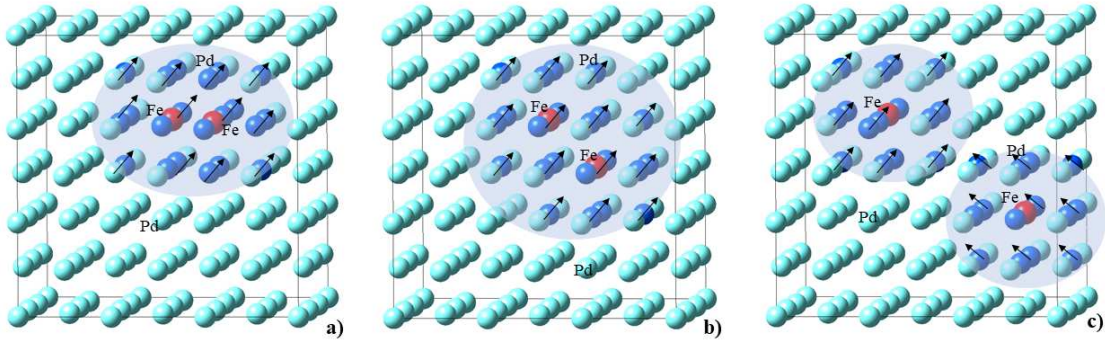}
	\caption{Magnetic clusters formed around Fe solute atom  impurities in the  3$\times$3$\times$3 Pd host matrix. Three configurations are shown: (a) Fe atoms are located close  to each other, (b) further away  and (c) at a distance.  Red spheres of atoms  correspond to Fe, light blue -- to Pd, blue -- to predominantly polarized Pd atoms. Transparent blue spheres outline magnetic clusters formed around magnetic impurities and black arrows denote one special case of magnetic moments orientation.}
	\label{cluster}
\end{figure}

With this finding, we may answer  the  question mentioned in the introduction section.
After implantation, impurities are located inhomogeneously: some Fe solute atoms are too close, others --  too distant. However, the annealing process used in experimental investigations leads to a more uniform redistribution of impurities in the host Pd-matrix with lower energy. Obviously, the system tends to have the  lowest energy, and this may happen during  annealing. 
This is illustrated in Fig.~\ref{cluster}, where three possible iron configurations are discussed. 
%Fig 5
%
%1
Closely located magnetic impurities (Fig.~\ref{cluster}\,a) form two overlapping  magnetic clusters of neighboring Pd atoms. This situation is energetically unfavorable and the system of Pd$_{0.98}$Fe$_{0.02}$  has a relatively low total magnetic moment $\approx$\,0.5\,$\mu_{B}$ (Fig.~\ref{vs_position}\,a) calculated per Fe atom.
%2
As soon as solute atoms move away from each other (during annealing, for instance) (Fig.~\ref{cluster}\,b), the total energy lowers reaching a minimum (Fig.~\ref{vs_position}\,c). This energy minimum corresponds to the maximum of magnetization ($\approx$\,8\,$\mu_{B}$ at \,0.2-0.35\,nm in Fig.~\ref{vs_position}\,a) and the solute atoms configuration depicted in Fig.~\ref{cluster}\,b.
%3
Finally, distant magnetic clusters (Fig.~\ref{cluster}\,c) weakly feel each other and may even have  opposite magnetic moment directions, which results in low spontaneous magnetization and higher energy.

\section{Conclusion}
In the present work, we have  demonstrated that calculations within the density functional theory can reproduce basic experimental results obtained for Pd$_{1-x}$Fe$_{x}$ alloys, in particular, we have shown that Fe solute  in Pd host matrix induce ferromagnetism, which extends to very low concentrations ($x=0.03$). The magnitude of the total magnetic moment is very large, about 8\,$\mu_{B}$ per Fe solute atom and it smoothly decreases with increasing impurity concentration.
%[как и выше, здесь не понятно, что подразумевается по термином induced magnetic moment, т,к. в рукописи нет его определения]

We also investigated the dependence of magnetization on the mutual positions of dissolved Fe  in the Pd host matrix. The aim was to answer the experimental question about the nature of the increase in $T_{C}$ and $\textit{M}$  upon annealing. We have  demonstrated that the magnitude of the total magnetic moment of Pd$_{0.98}$Fe$_{0.02}$ alloy depends on impurity positions, and, as a consequence, the magnetization increase during annealing relates to the impurity redistribution in the Pd matrix.

%Finally, in the last section we have also presented results for Co$^{+}$, Mn$^{+}$, Er$^{+}$ containing alloys. Co$^{+}$ impurities able to produce high magnetic moment of  $\approx$15\,$\mu_{B}$ per Co atom, whereas Mn$^{+}$ and Er$^{+}$ do not produce significant magnetisation as follows from our \textit{ab initio} calculations.

\section*{Acknowledgments}
This work was supported by RFBR Grant No. 20-02-00981. Computing resources were provided by Laboratory of Computer design of new materials in Kazan Federal University.

\end{document}